# SHADOWING EFFECTS ON ROUTING PROTOCOL OF MULTIHOP AD HOC NETWORKS

Md. Anwar Hossain, Mohammed Tarique and Rumana Islam

Faculty of Engineering, American International University-Bangladesh
*anwar94113@aiub.edu*

**ABSTRACT**

*Two-ray ground reflection model has been widely used as the propagation model to investigate the performance of an ad hoc network. But two-ray model is too simple to represent a real world network. A more realistic model namely shadowing propagation model has been used in this investigation. Under shadowing propagation model, a mobile node may receive a packet at a signal level that is below a required threshold level. This low signal level affects the routing protocol as well as the medium access control protocol of a network. An analytical model has been presented in this paper to investigate the shadowing effects on the network performance. The analytical model has been verified via simulation results. Simulation results show that the performance of a network becomes very poor if shadowing propagation model is used in compare to the simple two-ray model. Two solutions have also been proposed in this paper to overcome the effects of shadowing. One solution is a physical layer solution and the other one is a Medium Access Control (MAC) layer solution. Simulation results show that these two solutions reduce the shadowing effect and improve network performance.*

**KEYWORDS**

*Ad hoc networks, shadowing, link distance, probability, delivery ratio, signal-to-interference-plus-noise power ration (SINR), MAC layer, and transmission power.*

## 1. INTRODUCTION

Mobile Ad hoc Network (MANET) is a highly appealing means of providing network support to a group of people without the aid of any infrastructure support. MANET consists of a group of mobile nodes that may not be within the range of each other. Necessary controls and networking functions are performed by using of a distributed control algorithm. MANET is characterized as a multihop communication network unlike a cellular mobile network where each mobile node is connected to a nearest base station based on a single hop connection. In MANET, a mobile node forwards packet for other mobile nodes in addition to transmit its own packet. Dynamic topology is another important characteristic of MANET [1]. Mobile nodes may join or leave a network at any time. In order to maintain network connectivity under a dynamic topological condition an efficient routing protocol is very essential for MANET. Existing routing protocols like Destination Sequence Distance Vector (DSDV) [2], Dynamic Source Routing (DSR) [3] and Ad-hoc On-Demand Distance Vector (AODV) [4] consist of two main mechanisms namely route discovery and route maintenance. A mobile node discovers a route or a set of routes to a destination mobile node by using a route discovery mechanism. On the other hand, a mobile node detects any network topology change by using the route maintenance mechanism. While using any of these two mechanisms, a routing protocol relies on mobile radio channel. Mobile radio channel places a fundamental limitation on the performance of a MANET. The transmission path between a transmitter and a receiver can vary from simple line-of-sight to one that is severely obstructed by buildings, trees, road signs, mountains, and other objects. Hence mobile radio channel is extremely random unlike its wired counterpart. Different propagation models have been proposed in the literatures to predict the signal attenuation that occurs between two mobile nodes separated by a distance. The ground reflection model or two-ray propagation model is widely used in the test bed and also in the simulation model [10-17]. Two-ray propagation model assumes that there is a line-of-sight path and a ground reflected propagation path between a transmitter and a receiver. This model has been found to be reasonably accurate for predicting the large-scale signal strength over distances of several kilometers for mobile radio system. This model characterizes signal propagation in an isolated area with few reflectors such as rural road or highways. It is not typically a good channel model for a real world mobile communication system especially when the system is deployed in an urban area. Because this model does not consider the fact that the surrounding environment of a network is always changing. In a real world situation, the surrounding environmental cluster is always changing. This leads to a signal level that can vary vastly across a given distance. The short-term variations in the signal strength can be as high as 10-20 dB from the mean value of the signal. Measurements have shown that at any given distance, the path loss at a particular distance is random and distributed log-normally about a mean value [5]. The log-normal distribution describes the random shadowing effects which occur over a large number of measurements locations that have the same separation distance. This adverse effect of shadowing on the routing protocol has been investigated in this paper.

One of the first papers related to connectivity issues in wireless multihop network was [18]. The authors studied the percolation of a broadcast in a multihop radio network modeled by a spatial Poisson process. The effect of station density and transmission radius on the extent of broadcast percolation was examined. The results presented in this paper shows that in optimizing transmission radius as a function of communication performance measures, the choice of radius may be bounded from below by the need to maintain a desired network connectivity. Another early paper [19] addressed connectivity issues for mobile nodes that are randomly distributed according to a uniform probability distribution on a one-dimensional line segment. More recently another work [20] performed a fundamental study on the connectivity of uniformly distributed mobile nodes on a circular area. According to [20] mobile nodes should adjust transmission power to a level that is just enough to maintain connectivity in a network provided a mobile node cooperates with other mobile node to route packet. Further analytical investigations of the connectivity in bounded areas were made in [21- 22]. In [21] the authors analyzed the critical transmission range for connectivity in wireless ad hoc networks. The authors first considered the connectivity problem for stationary network and provided an upper and lower bound on the critical transmission range for one dimensional network. The authors evaluated the relationship between the critical transmission range and the minimum transmission range that ensured formation of a connected component containing a large fraction (i.e., 90%) of the nodes. The authors then extended this work to a mobility condition where mobile nodes were allowed to move during a time interval. In [22] the authors considered a *d*-dimensional region, with $1=d=3$ and they presented an analysis to determine the transmission range that ensure the resulting network is connected with a high probability. Based on the bounds of node density the authors concluded that, as compared to the deterministic case, a probabilistic solution to this range assignment problem achieves substantial energy savings. A framework for the calculation of stochastic connectivity properties of wireless multihop network has been presented in [23]. In fact, the connectivity problem has been solved for the general case of a k-connected network accounting for the robustness against node failures. These issues were studied for uniformly distributed nodel, Gaussian distributed nodes, and nodes that move according to the commonly used random waypoint mobility model. A large scale network with low node density has been investigated in [25]. The author studied the connectivity for both purely ad hoc network and hybrid network. In hybrid network, base stations were placed in a network. Mobile node communicates with other mobile node through the base stations. The authors obtained an analytical expression for the probability of connectivity in one dimensional network. They showed that bottlenecks are unavoidable in a low density sparse network. Most of the works mentioned so far assume idealized radio propagation model without considering fading and shadow effects. One of the earliest papers that considered fading and shadowing is [26]. The authors claim that many well designed protocols will fail simply because of fading and shadowing experienced in a realistic wireless environment. The authors have shown that fading and shadowing can have significant influence on network performance. They studied three different systems namely (1) a multichannel CDMA system, (2) a pure CDMA system, and (3) a contention based system. They also show that the multichannel CDMA system outperform the pure CDMA system as well as the contention based system under fading and shadowing environments. The connectivity of multihop radio networks in a log-normal shadow fading environment has been investigated in [27]. Assuming the mobile nodes have equal transmission capabilities and are randomly distributed according to a homogenous Poisson process, the authors provided a tight lower bound for the minimum node density that is necessary to obtain an almost surely connected subnetwork on a bounded area of given size. The authors also provided an insight into how fading affects the topology of multihop networks. The connectivity of a network from a layered perspective has been investigated in [28]. The authors first pointed out how the transmission range affects the end-to-end connection probability in a long-normal shadowing model and compared the results to theoretical bound and measurements in the path loss model. The authors then showed how connectivity issues behave in 802.11 and IP based networks if fading effects increases. The authors came up with an analytical model for the link probability in log-normal shadowing environments as a function of the number of nodes, network area, transmission range, path loss and shadowing deviation.

A probabilistic analysis of the shadowing effects on the signal level variations has also been presented in this paper. While investigating the impact of shadowing on the network performance, the delivery ratio has been considered as a performance metric. The delivery ratio is defined as the ratio between the number of packets received at a destination and the number of packets that was sent by a source to that destination.  Hence the packet delivery ratio indicates how many packets were lost in a network. An analytical model to estimate packet losses in a given network has been presented in this paper. This analytical model has also been verified via simulation results. In order to investigate the shadowing effects on a routing protocol, we selected Dynamic Source Routing (DSR)[3] protocol as the candidate. A brief description of the DSR protocol has been provided in the following section for the completeness of this work. The effects of shadowing on the DSR protocol have been explained in the section III. The shadowing has grave effects on a medium access control scheme. Since IEEE 802.11 MAC layer protocol has been used in this paper, section IV contains a brief description of IEEE 802.11 MAC layer protocol and section V shows the effects of shadowing on IEEE 802.11 MAC layer protocal. Using a derived probability density function based on the results published in [6], the mean value of the link distances has been derived in section VI. The probability that the received signal level will be greater than a threshold level has also been derived in the same section. Simulation models and results have been presented in section VII. Two solutions of shadowing problem have been presented in the same section. Finally, we conclude this paper in the last section.

## 2. THE DSR PROTOCOL

The DSR [3] protocol consists of two basic mechanisms: (1) route discovery, and (2) route maintenance. Route discovery is the mechanism by which a source node discovers a route to a destination. When a source node wants to send a data packet, it first looks into its route cache to find a route. If a source cannot find a route in its route cache, it initiates a route discovery mechanism by broadcasting a request packet to its neighbours. When a neighbour of a source receives a request packet, it first checks whether the request packet is intended for it or not. If a neighbour discovers that it is the destination, it sends a reply back to the source after copying the accumulated routing information contained in the route request packet into a route reply packet. If a neighbour discovers that it is not the destination of this request packet, it checks for a route in the route cache for that destination. If this neighbour is neither a destination nor it has a route in the route cache to that destination, it appends its address in the route request packet and re-broadcasts the route request packet to its neighbours. This process continues until a route request packet reaches at the destination node. Then the destination node replies all route requests that it receives. When a source node receives a route reply packet, it starts sending data packets using the route indicated in the reply packet. If multiple paths are discovered, it chooses a path that is the shortest one.

A typical route discovery mechanism of DSR protocol is illustrated in Figure 1. In this scenario, mobile node '1' is the source and mobile node '5' is the destination. Mobile node '1' does not have any route in its cache to this destination. Hence it initiates the route discovery mechanism by broadcasting a request packet. When some neighbouring nodes like mobile node '2' and '6' receive the request packet, they put their addresses in the request packet and re-broadcast that request message. Similarly mobile nodes '3','4' and '7' also rebroadcast that request packet

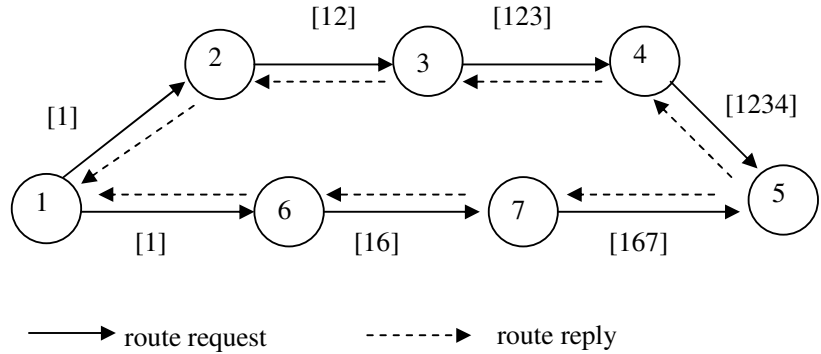

Figure1. Route discovery mechanism of DSR protocol until the destination node '5' receives it. The destination node '5' then replies back to the source node '1' by using a route reply packet. The route reply packet contains the information of the route that has been just discovered. After receiving the route reply packet, mobile node '1' records all the routes in its route cache. In this simple scenario, two routes namely '1-2-3-4-5' and '1-6-7-5' are discovered. According to the routing algorithm, mobile node '1' should select the shortest path between these two paths. In this case, the shortest path is '1-6-7-5'. After selecting this shortest path, the source node starts sending data packet by using the route '1-6-7-5'.

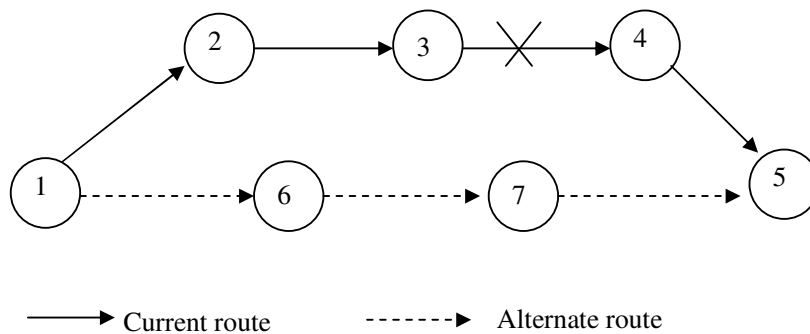

Figure 2. Route maintenance mechanism of DSR protocol

Route maintenance is the mechanism by which a node is able to detect any change in the network topology. When a node detects a 'broken' link, for example, by using missing MAC layer acknowledgments, it removes the link from its route cache and sends a route error message to each node that has sent packets over that link. A typical route maintenance mechanism is shown in Figure 2. Let us assume that the link between mobile node '3' and '4' is 'broken' due to battery exhaustion of mobile node '4'. Mobile node '3' detects that the mobile node '4' is unreachable by using a MAC layer mechanism. Mobile node '3' then creates a route error message and sends it to the source. A route error message contains the information of the faulty link (i.e., 3-4). After receiving the route error message, the source mobile node '1' marks the route '1-2-3-4-5' as invalid route and tries to find an alternative route from its route cache. Since an alternative route '1-6-7-5' is there in the route cache, the source mobile node should select that route and start using this new route.

## 3. EFFECTS OF SHADOWING ON ROUTING PROTOCOL

In the route discovery and route maintenance operation of the DSR protocol, it is assumed that the link between two nodes is stable and the variation of signal level only depends on the distance between them. That means a neighbouring node is always 'reachable' unless this neighbour is not out of battery or it has moved out of the reach. But shadowing effect assumes that the signal level can vary widely for a given distance between two nodes. This increases the probability that the signal level may go below a certain required level called a 'threshold' level. In this case, a receiving mobile node may not successfully receive a packet. Hence the following problems may arise:

(a) The route request packet may not reach all the neighbours. Hence there is a probability of an unsuccessful route discovery. That means the route request packet may not reach to a destination.

(b) The route discovery mechanism may not be an efficient one. The route discovery mechanism of the DSR protocol aims to discover as many paths as possible. The reason is that if one path fails, a source can selects an alternative path instantly. But this kind of flexibility may be lost under shadowing condition. Since only a few number of paths are discovered, there is a probability that a source may not find any other alternative route once a current route fails.

(c) Once some routes are discovered and a source mobile node starts sending data packets using one of the discovered routes, a neighbouring node may not receive that data packet because of the wide variation of signal. Hence a data packet may be lost at an intermediate node.

(d) The route maintenance operation of the DSR protocol also may not work properly because there is a probability that a route error message may be lost during its way to a source mobile node. If a source does not receive the route error message, it cannot detect a link breakage. The source continues sending data packet using the route that contains the broken link. All these data packets will be lost at the broken link.

The shadowing not only affects a routing protocol but also makes problems for a medium access control scheme. In all the simulation results presented in this paper, IEEE 802.11 has been chosen as the medium access protocol. A brief description of IEEE 802.11 MAC layer has been described in the following section.

## 4. IEEE 802.11 MAC LAYER

Like other IEEE 802.x protocol, the IEEE 802.11 protocol defines the Medium Access Control (MAC) and physical layers. The IEEE 802.11 MAC layer defines two different access methods namely Distribution Coordination Function (DCF) and Point Coordination Function (PCF). We will now only describe DCF based MAC protocol since PCF based MAC protocol cannot be used in an ad hoc network. The basic access mechanism of DCF based MAC protocol is basically a Carrier Sense Multiple Access with Collision Avoidance (CSMA/CA). The main mechanism of CSMA/CA protocol is as follows: a mobile node senses the medium before transmitting its packet. If it finds that the medium is free, it transmits its packet. But if the medium if busy, a mobile node defers its transmission to a later time which is chosen randomly. IEEE 802.11 protocol includes both physical carrier sensing and virtual carrier sensing.

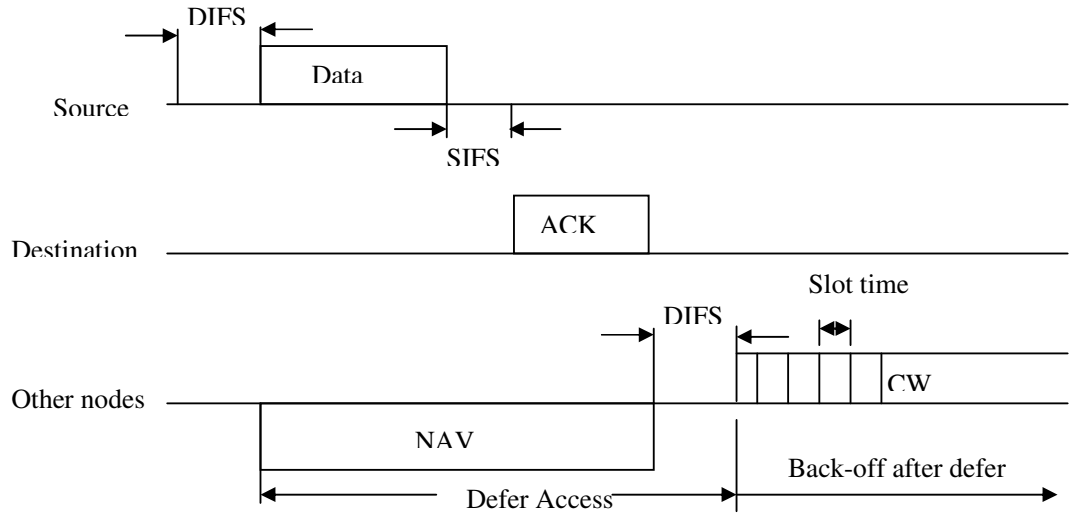

Figure 3. Carrier sensing in IEEE 802.11

Physical carrier sensing is used to detect other mobile nodes operating in the same network by analyzing all detected packets. It also helps to detect activities in the channel via relative signal strength from other sources. Virtual carrier sensing is performed by sending MPDU duration information in the header of Request-to-Send (RTS), Clear-to-Send (CTS), and data frames as shown in Figure 3. In addition to header information, payload and a 32-bits CRC, MPDU also contains a duration field. The duration field contains information of the time duration that will take this data or management frame to complete transmission. Other mobile stations in the same area use this information to adjust their Network Allocation Vector (NAV), which indicates the amount of time that must elapse until the current transmission session is completed and the channel can be sampled again for idle status. The channel is marked busy if either the physical or virtual carrier sensing mechanism indicates the channel is busy. Priority access to the wireless medium is controlled through the use of Inter-Frame Space (IFS) time intervals between the transmissions of frames. Two IFS intervals are specified in IEEE 802.11 standard (a) Short Inter-frame Space (SIFS) and, (b) DCF Inter-Frame Space (DIFS). The SIFS is the smallest IFS after a DIFS as shown in Figure 3. When a station senses the medium and finds it is idle, it has to wait for a DIFS period to sense the medium again. If the channel is still idle, the mobile station transmits its MPDU. Upon receiving the MPDU, a receiver checks the checksum to determine whether a packet was received correctly or not. If a packet is received correctly, a receiver waits for a SIFS time period and sends a positive acknowledgement (ACK) to the transmitting station to indicate that the transmission was successful. If an acknowledgement is not received within a given time period, another attempt is made to send a packet again. The number of times a source attempts to send a packet is determined by an important parameter named 'Long Retry Limit'. When the number of attempts exceeds this limit, a source discards a packet permanently. Since each mobile node has a limited buffer to store packet, this 'Long Retry Limit' parameter helps a mobile node to operate with a limited packet buffer size. In order to reduce collision and save channel bandwidth, Request-to-Send (RTS) and Clear-to-Send (CTS) are also used in IEEE 802.11 layer. RTS and CTS control frames are used by a station to reserve channel bandwidth prior to the transmission of MPDU. The timing diagram of RTS and CTS packets are shown in Figure 4. The RTS control frame is first transmitted by a source mobile. All other stations in a given area read the duration field and set their NAVs accordingly. The destination station responds to the RTS packet with a CTS packet after an SIFS idle period has elapsed. Mobile stations hearing the CTS packet look at the duration field and again update their NAV. Upon successful reception of the CTS, the source station is virtually assured that the medium is stable and reserved for successful transmission of the MPDU. In this way, the mobile stations update their NAVs based on the information contents of RTS and CTS packets, which helps to combat 'hidden terminal' problem. The collision avoidance portion of CSMA/CA is performed through a random back-off procedure. If a mobile station with a frame to transmit senses the channel and finds the medium is busy, then the mobile node waits for a DIFS period. At the end of the DIFS period, the mobile station computes a random period of time called back-off period. In IEEE 802.11, time is slotted in time periods that corresponds to a time unit called slot time. The random back-off time is an integer value that corresponds to a number of time slots. Initially, the mobile station computes a back-off time in the range of 0-31. This period is called the contention window as shown in

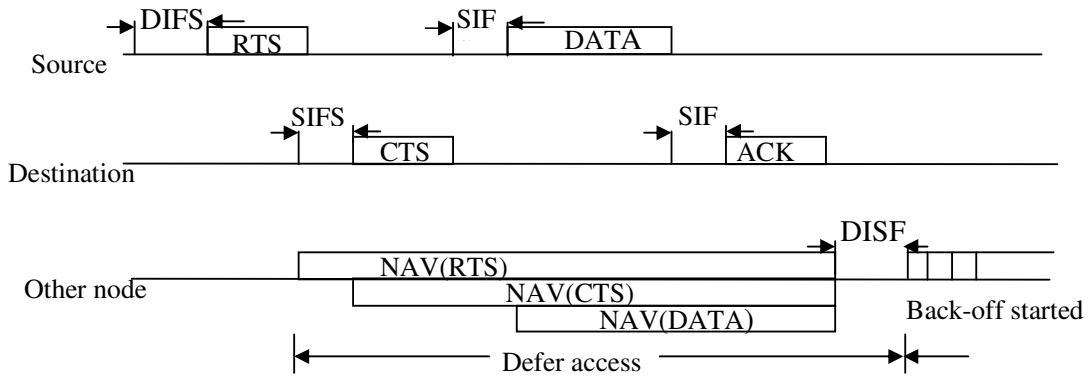

Figure 4. Transmission of MPDU with RTS and CTS

figure 4. After this contention period, a mobile node again senses the medium. If it finds the medium busy again, it goes for longer contention period. On the other hand, if it finds the medium free, it waits for DIFS period to make sure no other mobile node is transmitting. If no other mobile node transmits during that period, a mobile node starts transmitting its own packet.

## 5. EFFECTS OF SHADOWING ON MAC LAYER

The CSMA/CA based multi-access technique assumes that the medium is an intermittent synchronous multi-access bit pipe on which idle periods can be distinguished from packet transmission periods. If nodes can detect idle periods quickly, it is reasonable to terminate idle periods quickly and to allow nodes to initiate packet transmission after such idle detections. This is the philosophy of a CSMA/CA based multiple access technique. Shadowing effects may put hurdle on the normal operation of IEEE 802.11 MAC layer operation in the following ways: (a) as mentioned in the earlier section that IEEE 802.11 carrier sensing is performed at both the air interface, which is referred to as physical carrier sensing and at the MAC sub-layer referred to as virtual carrier sensing. Physical sensing may not work properly due to shadowing because signal level may go below a threshold level so that it cannot be detected. Hence both physical carrier sensing and virtual carrier sensing may not work properly, (b) a source station performs virtual carrier sensing by sending MPDU duration information in the header of RTS, CTS and data packets. Stations in a given area use this information in the duration field to adjust their Network Allocation Vector (NAV). The NAV indicates the amount of time that must elapse until the current transmission session is complete and the channel can be sampled again for idle status. The stations in a given area may not be able to recover the bit duration information from the packet because of the poor signal level arisen from the shadowing effects. Hence the adjustment of NAV cannot work properly, (c) The RTS and CTS frame exchanging between a source station and a destination station may not be successful due to variation of signal level. If the destination station does not receive RTS packet due to signal variation, it does not reply that RTS by sending a CTS packet. Since a source station cannot send a data packet unless it receives a CTS packet from the destination, a source station has to keep the data packet in the buffer for longer period of time. Similarly, if the CTS packet is not successfully received by a source station due to signal variation, a source station also delays its transmission. Hence unsuccessful reception of RTS and CTS packets can cause unnecessary delaying of a packet transmission, (d) according to IEEE 802.11 after receiving a data packet, a source station sends an acknowledgement. A destination station may not receive a data packet successfully due to variation of the signal strength. Hence a source station has to resend a data packet several times. These redundant packets will occupy channel unnecessarily and waste scarce channel bandwidth. On the other hand, a destination may successfully receive a packet, but the acknowledgement packet sent by that destination may be lost due to shadowing effects. Since a source node does not receive an ACK packet, it keeps sending the data packet repeatedly. According to IEEE 802.11 MAC layer, a source has the opportunity to resend a packet for seven times after that it will assume that the destination is unreachable and it drops the packet. Hence there will be a large number of packet losses in a network due to unsuccessful reception of an ACK.

## 6. DERIVATION OF EXPRESSION

In a uniform random network scenario, the location of a mobile node is determined according to uniform random variables. That means the location of a mobile node is determined by $x$ co-ordinate and $y$ co-ordinate that are two uniform random variables between 0 to A and 0 to B respectively, where A and B are the length and width of a network. To determine the average link distance between a given source-destination pair, we rely on the link distribution model presented in [6]. It is shown therein that when a number of mobile nodes are uniformly distributed over a rectangular area the probability density function $p_d(\gamma = \xi D_1)$ of the link distances between any two mobile nodes can be expressed as follows:

$$p_d(\gamma = \xi D_1) = \frac{1}{D_1} \begin{cases} \zeta\xi[2\zeta\xi^2 - 4\xi(1+\zeta) + 2\pi], & 0 \leq \xi < 1 \\ 4\zeta\xi\sqrt{\xi^2 - 1} - 2\zeta\xi(2\xi + \zeta) + \\ 4\zeta\xi\sin^{-1}(1/\xi), & 1 \leq \xi < \zeta^{-1} \\ 4\zeta\xi\sqrt{\xi^2 - 1} + 4\zeta^2\xi\sqrt{\xi^2 - \zeta^{-2}} - \\ 2\xi(\zeta^2\xi^2 + 1 + \zeta^2) + 4\zeta\xi\{\sin^{-1}(1/\xi) - \cos^{-1}(1/\zeta\xi)\}, & \zeta^{-1} \leq \xi < \sqrt{1+\zeta^{-2}} \\ 0, & otherwise \end{cases} \quad (1)$$

,where $\zeta = \dfrac{D_1}{D_2}$ is the shape parameter of a rectangular area, $D_1$ is the width and $D_2$ is the length of the rectangular area, $\xi = \gamma D_1$ is given by $0 < \gamma \leq \sqrt{D_1^2 + D_2^2}$. The mean value of the link distance is defined by $E[\xi] = \int \xi p_d(\xi D_1) d\xi$. Figure 5 depicts the variation of the mean link distance and number of hops between a source and a destination node. It is shown in this figure that the link distance variation is almost linear with respect to the network area of operation. If the shadowing effect is ignored, the received power level at any mobile node in an ad hoc network depends on the link distance between two mobile nodes. Since the mobile node has limited transmission power, a packet should travel multiple hops from a source to a destination. If the transmission range corresponding to a transmission power is fixed and it is denoted by R, then the number of hops a packet should travel is given by $H = \left\lceil \frac{E[\xi]}{R} \right\rceil$. If we do not consider the shadowing effects, then the quality of each link should remain constant during the operation of the network. But this is not a valid assumption under a shadowing condition. Shadowing causes a wide range variation of the received signal. It increases the probability that the received signal goes below a threshold level. Hence a packet may not be correctly received by a receiver. The analytical model presented in this section is based on the following assumptions: (1) all link distances are uniform within a rectangular service area, (2) the characteristic of the channel is almost same over the whole rectangular service area (i.e., the whole network has shadowing effect), (3) the node density (i.e., number of nodes per square meter) is kept constant when network size is varied, and (4) once a packet is successfully received, the receiving mobile node re-transmits that packet at the same power level at which the previous node has transmitted the same packet to itself. Shadowing effect states that at any given distance $d$ from a transmitter, the path loss $PL(d)$ at a particular location is random and distributed log-normally (normal in dB) about the mean distance-dependent value [5]. Therefore, path loss and received power at $d$ meter from the transmitter is modelled as follows:

$$PL(d)\,dB = \overline{PL}(d_0) + 10\,n\,\log(\frac{d}{d_0}) + X_\sigma \qquad (2)$$

where $\overline{PL}(d_0)$ is the average path loss at a reference distance $d_0$ and is given by $\overline{PL}(d_0) = -10\log_{10}(\lambda^2/(4\pi d_0)^2)$ [7], $\lambda$ is the wave length of the signal, $n$ is the path loss exponent, $X_\sigma$ is a Gaussian random variable with a mean value of $0$. If this path loss model is used, the received power $P_r(d)$ at distance $d$ from a transmitter is given by

$$P_r(d)\,dB = P_t - [\overline{PL}(d_0) + 10\,n\,\log(\frac{d}{d_0}) + X_\sigma] \qquad (3)$$

where $P_t$ is the transmission power. The probability density function of the path loss due to shadowing effects can be expressed as [8]

$$p_{PL(d)}(x) = \frac{1}{\sqrt{2\pi}\sigma} \exp\left(-\frac{(x - \overline{PL}(d_0) - 10n\log 10(\frac{d}{d_0}))^2}{2\sigma^2}\right) \quad (4)$$

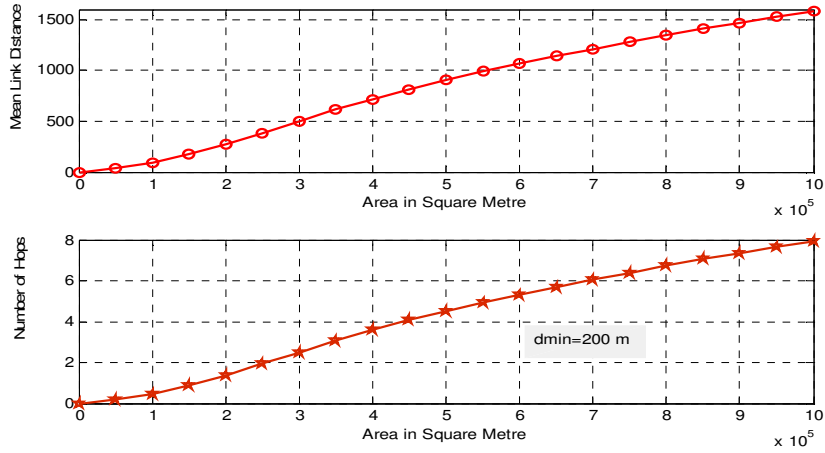

Figure 5. Mean link distance and number of hops for a rectangular service area

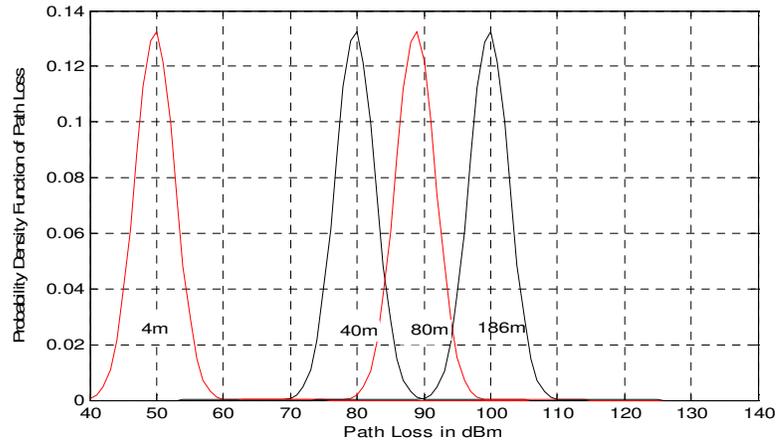

Figure 6. Comparison results of probability density functions of path loss at $d = 4$, 40, 80, and 186m

The probability density functions for path loss using Equation (4) is a Gaussian log-normal distribution. The distance dependent mean values of path losses at $d = 4, 40, 80,$ and 186 meter far from the receiver is 50, 80, 88, and 100 dBm respectively shown in the Figure 6.

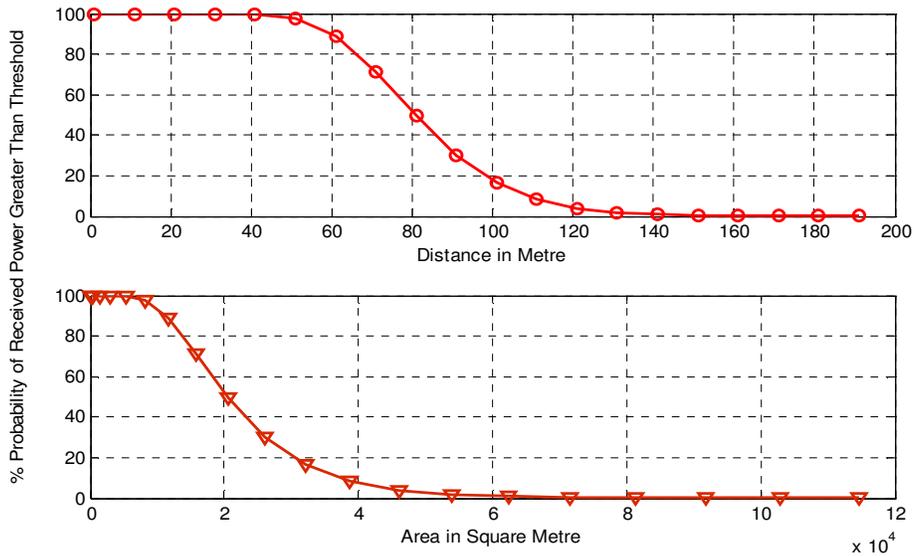

Figure 7. Probability of received power greater than threshold power level with respect to distance $d$ and circular area $\pi d^2$

Based on the analysis presented in [5], the probability that the received signal level will be greater than the threshold power $P_{th}$ can be expressed as

$$P_r(P_r(d) > P_{th}) = \frac{1}{2} - \frac{1}{2} erf(\frac{P_{th} - (P_t - (\bar{PL}(d_0) + 10n\log(\frac{d}{d_0})))}{\sigma\sqrt{2}}) \qquad (5)$$

where $\sigma$ is the standard deviation (in dB), $P_{th}$ is the threshold power and $erf(\cdot)$ denotes error function. Figure 7 shows the probability of received power greater than the threshold power $P_{th}$ is plotted with respect to distance $d$ and circular area $\pi d^2$ considering the shadowing effect in microcellular environment. The probability of the percentage of the network area where the received signal is greater than the threshold level $P_{th}$, is denoted by $P_r(P_r(d) > P_{th})$ is plotted in Figure 7. Figure 7 shows that if the distance between two nodes is less than 40 meter, the shadowing effect is not very significant. There is almost 100% probability that the received signal will be above a threshold level. But the probability $P_r(P_r(d) > P_{th})$ decreases exponentially as the distance between two nodes becomes greater than 40 meter and it becomes eventually 5% when the distance between two nodes is around 165-175 meter. From this observation, it can be concluded that although the shadowing effect is not significant for a small network, but the shadowing effect will be more severe for a large network. When the average link distance is more than 165 meter, there will be almost 95% packet loss in a network.

## 7. SIMULATION MODEL AND RESULTS

To investigate the effects of shadowing on the performance of an ad hoc network, a network consists of 30 mobile nodes was created and tested via Network Simulator (NS-2) [9]. These nodes were placed randomly over an area of $400m \times 300m$ area. Five connections were randomly set up in the network. While setting up each connection of the DSR protocol was used as the routing algorithm. Once a connection is set-up, Constant Bit Rate (CBR) agent was used to generate packets. Each CBR connection started at random period of time. Once a CBR connection started, it continued generating packet till the end of the simulation. The packet generation rate was 1 packet per second. Each simulation was tested for 250 seconds simulation time. IEEE 802.11 MAC layer was used as the MAC layer. The network size was then increased to $400m \times 400m$, $500m \times 400m$, $500m \times 500m$, $600m \times 500m$, $600m \times 600m$, $700m \times 600m,$ and $700m \times 700m$ by keeping the node density constant so that the network connectivity is not affected. That means there are 40, 50, 62, 75, 90, 105 and 122 mobile nodes were deployed over the network area when the network sizes were $400m \times 400m,$ $500m \times 400m,$ $500m \times 500m,$ $600m \times 500m,$ $600m \times 600m,$ $700m \times 600m,$ and $700m \times 700m$ respectively. The total number of data packets generated in the network at the source mobile nodes and the total number of data packets delivered to the destination mobile nodes were monitored during each simulation. The delivery ratio is the ratio defined by $\gamma = \dfrac{N_{recvd}}{N_{sent}}$. For a given area, ten different topologies were created and tested by using different seeds. Ten simulation results were then averaged. The other simulation parameters are shown in Table I. The transmission range is 250 meter with the given transmission power level of 24.50 dBm and the threshold power of -64.38 dBm. These are the default values of these two parameters in Network Simulator (NS-2) [9].

Table 1: Simulation parameters

| Parameters | Values |
| --- | --- |
| Transmitting Power $P_t$ | 24.50 dBm |
| Threshold Power $P_{th}$ | -64.38 dBm |
| Transmitting Antenna Gain $G_t$ | 1 |
| Transmitting Antenna Height $h_t$ | 1 m |
| Receiving Antenna Gain $G_r$ | 1 |
| Receiving Antenna Height $h_r$ | 1 m |
| Shadow Standard Deviation $\sigma$ | 3 dB |
| Close Reference Distance $d_0$ | 1 m |
| Path Loss Exponent $n$ | 3.0 |
| Carrier Frequency $f$ | 914 MHz |
| Propagation Models | Shadowing and Two-ray model |

The two-ray model has been used as the propagation model. As mentioned in the previous section that two-ray model is too simple to represent a real world scenario. The two-ray reflection model assumes that there are two paths between a source and a destination. One path is the line-of-sight path and the other one is the reflected path from the ground. The variation of the signal strength follows the following rule [5]

$$P_r = P_t G_t G_r \frac{h_t^2 h_r^2}{d^4} \qquad (6)$$

where $G_t$ and $G_r$ are the transmitting antenna gain and receiving antenna gain, $h_t$ and $h_r$ are the transmitting and receiving antenna heights, $P_t$ is the transmitting power and $P_r$ is the receiving power, and $d$ is the distance between the transmitter and the receiver. Based on the two-ray model the received signal variation is illustrated in the upper graph of Figure 8. The figure shows that the received power level remains above a threshold power level $P_{th}$ unless the distance between a transmitter and a receiver is 160 meter. After that the received signal level goes below a threshold level. Hence there will be packet losses in the network if a network is large enough to have average link distance greater than 160 meter. Since each mobile node uses the same power level to transmit all kinds of packets (i.e., route discovery, route maintenance and data packets), the link up to 160 meter is considered stable in this case. That means once a route is discovered, the qualities of all links lying along that route do not change over the time. Hence there is almost no packet loss. The lower graph of Figure 8 also depicts the simulation results. It shows that the delivery ratio for two-ray model is almost 100%. That means a negligible number of packets has been lost. This figure shows that the delivery ratio is 100% under different network size. Because the average link distance is those simulated networks is less than 160 meter. The above mentioned simulations were repeated with shadowing propagation model by keeping other parameters mentioned in Table I same. The simulation results with that of the probabilistic analytic model are shown in the Figure 9. It is depicted that the delivery ratio is almost 100% when network area is less than or equal to $1.6 \times 10^5$ square meter. After that the delivery ratio decreases exponentially as the network gets larger. The delivery ratio decreases exponentially to almost 15% when the network size was $700m \times 700m$. This exponential drop in the delivery ratio complies with the theoretical model derived in section VI where it was shown that the probability of the received power is greater than the threshold power level. In order to successfully receive a packet, the received signal level should be greater than the threshold value defined by the parameter (i.e. $P_{th}$), which is listed in Table 1. The decision of successfully received a packet is determined by the probability defined as $P_r(P_r(d) > P_{th})$ since successfully received packets solely depend on the received power that is a random variable for shadowing model. This probability is directly proportional to the delivery ratio within the same rectangular area where link distance is almost linearly increased with the rectangular area, i.e.,

$$\text{Delivery ratio} = k\ P_r(P_r(d) > P_{th}) \qquad (7)$$

where $k$ is a scaling factor constant. Both the theoretical and simulation results of shadowing model are shown in Figure 9. In addition to these results, simulation result of two-ray model is also plotted in the same graph to show the difference between the network performance under two-ray model and shadowing model. Two methods that can reduce the shadowing effects suggested in this paper are: (1) by increasing the transmission power, and (2) by increasing the retry-limit of MAC layer protocol. The first solution is a physical layer solution. This solution of reducing shadowing effects is to increase the transmission power level.

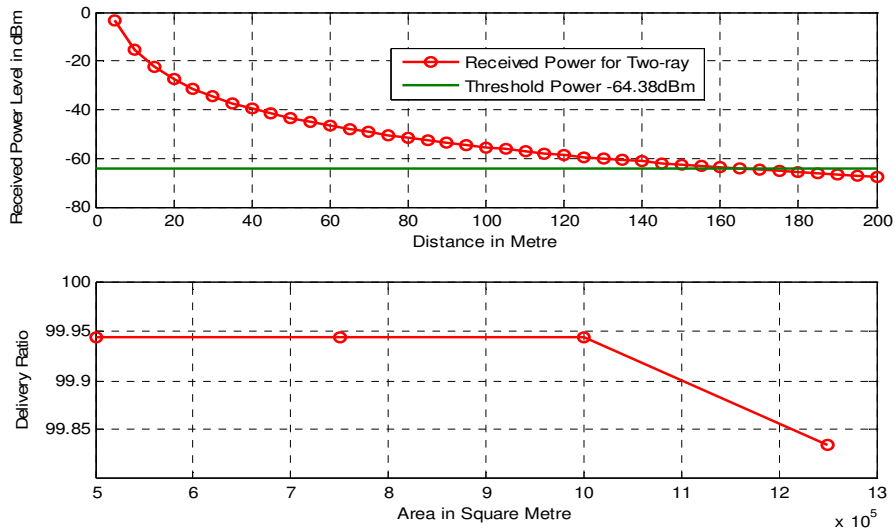

Figure 8. Delivery ratio of simplest two-ray channel model with respect to rectangular area

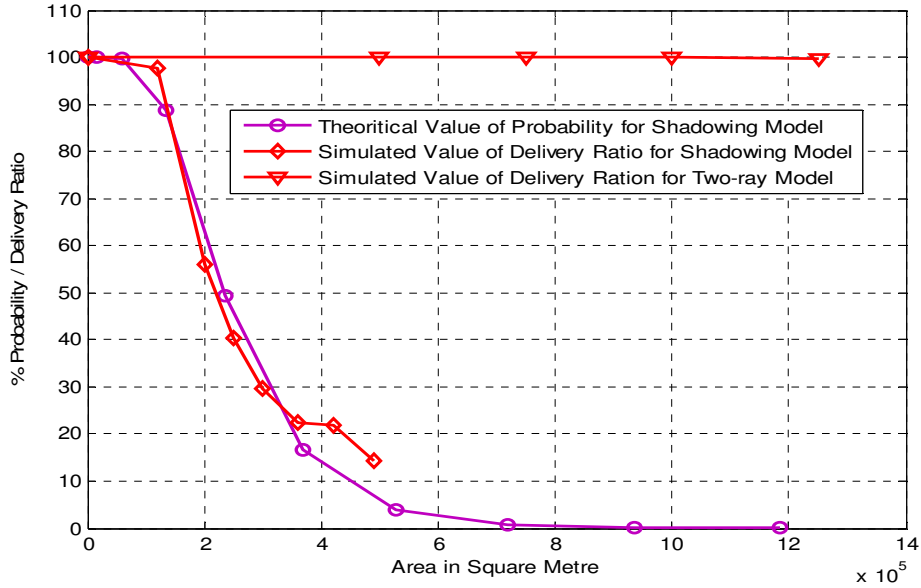

Figure 9. Comparison of simulated delivery ratios of two-ray channel model and shadowing model and theoretical probabilistic analytic result of shadowing model

As shown in Equation (3) that the received signal level at a given distance will increase if the transmission power $P_t$ is increased. To investigate how a higher transmission power can reduce the shadowing effect the previous simulations have been repeated here. But the transmission power is increased to 0.58432 watts (or 27.67 dBm). This transmission range of a mobile for this new transmission power is 300 meter (if two-ray model is used). The network performance in terms of delivery ratio under a shadowing condition but with the increase in transmission power is shown in Figure 10 labelled as high transmission power. The figure shows that the delivery ratio can be improved if a higher transmission power is used. For example, when the network area is $500m \times 500m$, the delivery ratios are 40% and 55% for the transmission power of 24.5 dBm and 27.67 dBm respectively. But the improvement in delivery ratio is less when the network size is small. The higher transmission power level not only increases the probability of improving the signal level. But it also helps to reduce the number of hops that a packet travels from a source to a destination. Hence there is less probability of packet loss. Although the higher transmission power improves delivery ratio, it may not be a good choice to increase the transmission power to reduce shadowing effect. High transmission power will increase interference level in a network. Another alternative solution to reduce the shadowing effect is to modify the MAC layer protocol. As mentioned previously that an important parameter of IEEE 802.11 MAC layer is 'Long Retry Limit'. By default the 'Long Retry Limit' is set to 7. That means after 7 attempts, a mobile node discards a packet permanently by assuming that the next hop is unreachable. But under a shadowing condition 7 attempts are not enough. A mobile node may not successfully receive a packet from its previous hop due to poor signal level in 7 attempts. In order to increase the probability of successfully receiving a packet the 'Long Retry Limit' was increased to a higher value of 12. That means a mobile node has 5 additional attempts to send a packet successfully to its next hop. The simulation result for this MAC layer solution is shown in shown in Figure 10 labelled as 'MAC Layer'. This figure shows that delivery ratio is improved compared to that of a shadowing. Although the amount of improvement in MAC layer solution is less in compare to its higher transmission power counter part, but still the packet loss can be reduced in a network. The figure shows that an average almost of 10% packet loss can be reduced in a network if MAC layer is modified as mentioned above.

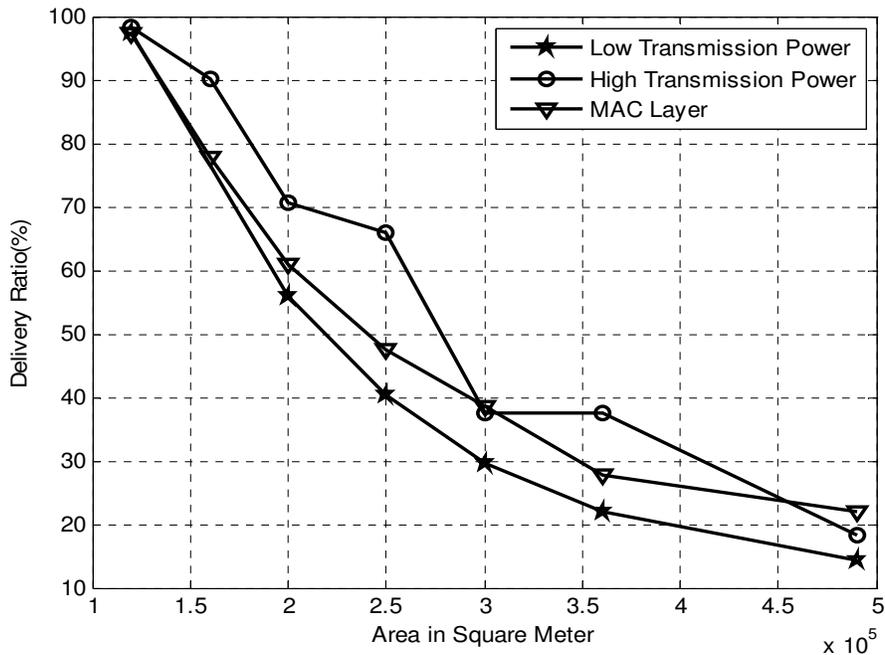

Figure 10. Comparison results of delivery ratio in shadowing environment using different Techniques

## 8. CONCLUSIONS

In this paper, the shadowing effects on the performance of an ad hoc network have been investigated. Although two-ray model is widely used as a propagation model in ad hoc network simulation, but this model does not represent a real world network propagation model because the surrounding environment of a network is always changing. It is shown that the performance of a network deteriorate very quickly if the shadowing effects are taken into account. The main reasons for this degraded performance resulted from the fact that there will be a large variation of the received signal level for a given link distance. Hence a packet (routing packet or MAC packet) may not be received successfully by a mobile node due to poor signal level. This causes problem to the normal operations of a routing protocol as well as the MAC protocol. One of the solutions of reducing showing effect suggested in this paper is to increase the transmission power level. The simulation result shows that the delivery ratio can be improved by almost 40% on average if the transmission power is increased from 24.5 dBm to 27.67 dBm. But higher transmission power increases the interference level in a network. Another alternative MAC solution has been suggested in this paper. This alternative solution is based on a modification of MAC layer protocol parameter. In this solution the number of packet transmission attempts was increased. The simulation result shows that this solution also improves the network performance.

**Authors**

Md. Anwar Hossain received the B.Sc. degree in Electrical and Electronic Engineering from Rajshahi University of Engineering & Technology, RUET in 2001 and M.Sc. Engineering degree in Information and Communication Technology from Asian Institute of Technology, AIT in 2006. He had worked as a lecturer in Electrical and Electronic Engineering Department at Rajshahi University of Engineering & Technology, RUET. Recently, he is working as an Assistant Professor at American International University-Bangladesh. His research interests are in Ad hoc networks, CDMA, OFDM and MIMO-OFDM Systems.

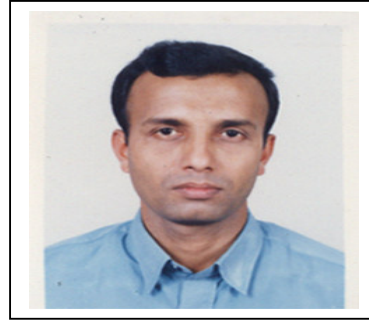

Mohammed Tarique received B.Sc. degree in Electrical and Electronics from Bangladesh University of Engineering and Technology (BUET) in 1992. He has worked in Beximco corporation for 6 years as a Senior Engineer. He received his Master of Business Administration (MBA) degree from the Institute of Business Administration (IBA), Dhaka. He also received his Master of Science (MS) degree from Lamar University, Texas, USA and Ph.D. degree from University of Windsor, Ontario, Canada. He is currently working in American International University-Bangladesh as an assistant Professor. His research interests are in wireless communication, sensor network, ad hoc network and mesh networks. He has presented his research works in the conferences held in Europe, USA and Canada. He has published several papers in the top tier journals.

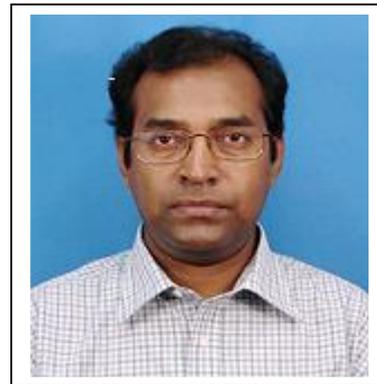

Rumana Islam completed her Master of Science (M.Sc.) in Biomedical Engineering from Wayne State University, Michigan, USA in 2005 and her B.Sc. Engineering degree in Electrical and Electronic Engineering from Bangladesh University of Engineering and Technology (BUET) in 1995. She is currently with the Department of Electrical and Electronic Engineering of American International University-Bangladesh. She has 7 years of professional experience in reputed public and private sectors. Her research interests are in Biomedical sensor design, sensor networks, ad hoc networks and Signal Processing.

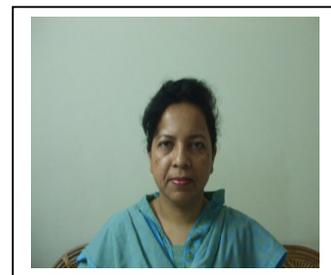